\documentclass[prd,nofootinbib,floats,superscriptaddress,eqsecnum,tightenlines,preprintnumbers,11pt]{revtex4-1}

\usepackage{hyperref}
\usepackage{graphicx}
\usepackage{amsmath,amssymb,amsfonts,amsthm,latexsym,stmaryrd}
\usepackage{marginnote}
\usepackage{color}

\def\be{\begin{equation}}
\def\ee{\end{equation}}
\def\beq{\begin{equation}}
\def\eeq{\end{equation}}
\newcommand{\bea}{\begin{eqnarray}}
\newcommand{\eea}{\end{eqnarray}}
\def\bi{\begin{itemize}}
\def\ei{\end{itemize}}
\def\ba{\begin{array}}
\def\ea{\end{array}}
\def\bfig{\begin{figure}}
\def\efig{\end{figure}}

\newcommand\cS{{\mathcal{S}}}

\newcommand\cR{{\mathcal{R}}}

\begin{document}

\preprint{YITP-20-81, IPMU20-0067}

\title{On Rotating Black Holes in DHOST Theories}

\author{Jibril Ben Achour}
\affiliation{Center for Gravitational Physics, Yukawa Institute for Theoretical Physics, Kyoto University, 606-8502, Kyoto, Japan}
\author{Hongguang Liu}
\affiliation{Institut fur Quantengravitation,  Universitat Erlangen-Nurnberg,  Staudtstr.  7/B2,  91058 Erlangen,  Germany}
\affiliation{Center for Quantum Computing, Pengcheng Laboratory, Shenzhen 518066, China}
\author{\\Hayato Motohashi}
\affiliation{Division of Liberal Arts, Kogakuin University, 2665-1 Nakano-machi, Hachioji, Tokyo, 192-0015, Japan}
\author{Shinji Mukohyama}
\affiliation{Center for Gravitational Physics, Yukawa Institute for Theoretical Physics, Kyoto University, 606-8502, Kyoto, Japan}
\affiliation{Kavli  Institute  for  the  Physics  and  Mathematics  of  the  Universe  (WPI), \\
The  University  of  Tokyo  Institutes  for  Advanced  Study, \\
The  University  of  Tokyo,  Kashiwa,  Chiba  277-8583,  Japan}
\author{Karim Noui}
\affiliation{Institut Denis Poisson (UMR 7013), Universit\'e de Tours, Universit\'e d'Orl\'eans, Parc de Grandmont, 37200 Tours, France}
\affiliation{Laboratoire Astroparticule et Cosmologie, CNRS, Universit\'e Paris Diderot Paris 7, 75013 Paris, France}

\date{\today}

\begin{abstract}
Using the disformal solution-generating method, we construct new axisymmetric solutions in Degenerate Higher Order Scalar Tensor (DHOST) theories. 
The method  consists in first  considering  a  ``seed'' known solution  in DHOST theories
and then performing 
a disformal transformation of the metric to obtain a new solution. In vacuum, the two solutions are equivalent but they become physically inequivalent when one considers coupling to matter.
In that way, we ``disform'' the stealth Kerr black hole solution and we obtain a first analytic rotating non-stealth  solution in DHOST theories, while the associated scalar field is time-dependent with a constant kinetic density. 
The new solution is characterized by three parameters: the mass, the spin and the disformal parameter which encodes the deviation with respect to the Kerr geometry. 
We explore some  geometrical properties of the novel disformed Kerr geometry which is no more Ricci flat, has the same singularity as the Kerr metric, admits an ergoregion, and is asymptotically flat. Moreover, the hidden symmetry of the Kerr solution is broken, providing an example of a non-circular geometry in a higher order theory of gravity.
We also discuss  geodesic motions and compute its (disformed) null directions which are interesting tools to understand the causal structure of the geometry. 
In addition, to illustrate again the potentiality of the disformal solution-generating method, we present another axisymmetric solution for DHOST theories obtained from a disformal transformation of the generalized Kerr solution of Einstein-Scalar gravity.
\end{abstract}

\maketitle
\tableofcontents

\section{Introduction}
Finding exact solutions for compact objects (stars, black holes or more exotic structures) 
in modified theories of gravity is in general a very difficult problem. Not only there exist many ``no-go'' theorems, extending the no-hair theorem of General Relativity (GR) which kill the hope of having new solutions, but also the equations of modified theories of gravity are, most of the time, much more complicated than  the Einstein equations, even for symmetry reduced space-times. Nevertheless, it was possible to
circumvent these difficulties and new black hole (analytical and numerical) solutions have emerged these last years, in particular, in the context of Degenerate Higher Order Scalar Tensor (DHOST) theories \cite{Langlois:2015cwa,Langlois:2015skt,Achour:2016rkg,Motohashi:2016ftl,BenAchour:2016fzp,Crisostomi:2016tcp,Crisostomi:2016czh} which provide the most general class of viable scalar-tensor theories to date evading the Ostrogradsky ghosts~\cite{Woodard:2015zca,Motohashi:2014opa,Motohashi:2020psc} 
(see also \cite{DeFelice:2018ewo} where DHOST theories have been extended to the so-called U-DHOST theories, and \cite{Motohashi:2020wxj} for further generalization). 
However, most of these solutions are stealth\footnote{Stealth black hole solutions were first introduced in the context of three dimensional gravity in \cite{AyonBeato:2004ig}. } spherically symmetric black holes \cite{Babichev:2016rlq,Babichev:2017guv,Lehebel:2017fag,Chagoya:2018lmv, BenAchour:2018dap,Motohashi:2019sen,Takahashi:2019oxz,Minamitsuji:2019shy,BenAchour:2019fdf,Minamitsuji:2019tet}. (See also \cite{Mukohyama:2005rw} for an earlier stealth solution in the context of ghost condensate~\cite{ArkaniHamed:2003uy}.)
The first rotating stealth solution was discussed in \cite{Babichev:2017guv} in the context of beyond Horndeski theories, and only recently, a more general stealth Kerr black hole solution was obtained in \cite{Charmousis:2019vnf} in the class of DHOST theories where gravitational waves propagate at the speed of light. 
On the other hand, the existence condition for any GR solutions with matter component minimally coupled to gravity was derived from a covariant analysis in \cite{Motohashi:2018wdq,Takahashi:2020hso} where the Kerr-Newman-de Sitter solution in quadratic DHOST theories was obtained.  
Moreover, solution-generating method such as the Kerr-Schild procedure have newly been extended to DHOST theories opening new avenues to derive interesting solutions \cite{Babichev:2020qpr}. Finally, a non-stealth numerical rotating black hole solution has also been constructed recently in \cite{VanAelst:2019kku}. 
Such rotating solutions represent an important step forward in order to investigate rotating compact objects in modified gravity. 
However, up to now,  analytic solutions describing a rotating compact object in DHOST theories \cite{Babichev:2017guv, Charmousis:2019vnf, Takahashi:2020hso} correspond to a stealth Kerr geometry and deviations from GR can only show up at the level of perturbations \cite{Charmousis:2019fre}. 
Therefore it would be interesting to construct new exact solutions beyond the stealth sector in order to obtain analytic solutions which exhibit deviations from GR already at the background level. The goal of this work is to present such a construction using the disformal solution-generating method discussed first in the context of DHOST theories in \cite{BenAchour:2019fdf}\footnote{See also \cite{Filippini:2017kov} for previous use of disformal transformation to construct a rotating black hole solution in vector-tensor theories.}. 
 
This method takes advantage of the properties of DHOST theories under disformal transformations of the metric, which were introduced in \cite{Bekenstein:1992pj}, to construct new exact solutions to the DHOST field equations from a known ``seed'' solution. 
As the class of DHOST theories is stable under disformal transformations, any known solution $(\tilde{g}_{\mu\nu},\phi)$ of a theory associated to the action $\tilde{S}[\tilde{g}_{\mu\nu},\phi]$ enables us to 
construct a ``new'' solution $(g_{\mu\nu},\phi)$ of a ``new'' theory whose action ${S}[{g}_{\mu\nu},\phi]$ is related to the previous one by,
\bea
\label{Intro:disformal}
S[g_{\mu\nu},\phi] = \tilde{S}[\tilde{g}_{\mu\nu},\phi] \, , \qquad
\tilde{g}_{\mu\nu} = A(\phi,X) g_{\mu\nu} + B(\phi,X) \phi_\mu \phi_\nu \, ,
\eea
where we are using the standard notations $\phi_\mu \equiv \partial_\mu \phi$ for the partial derivative of $\phi$ and $X=g^{\mu\nu} \phi_\mu \phi_\nu$
for its kinetic density. The functions $A$ and $B$ are, a priori, free but we will restrict our study to the cases where the
disformal transformation is invertible and where the two metrics are non-degenerate.  Although the construction of new exact solutions appears to be straightforward using this procedure, it is worth emphasizing that the obtained solution is not physically equivalent to the seed one in general. 
Indeed, in vacuum, the two actions $S$ and $\tilde S$
are equivalent in the sense that they have equivalent spaces of solutions provided that the disformal transformation is invertible. 
However, in the presence of matter minimally coupled to the metric (either $g_{\mu\nu}$ for the action $S$ or $\tilde{g}_{\mu\nu}$ for the disformed action $\tilde{S}$),
they become inequivalent. 
Hence, a particle falling into a black hole solution of the theory $S$ will have, in general, 
a different trajectory from that of a particle falling into the ``disformed black hole''  of  the theory $\tilde S$ (see \cite{Deffayet:2020ypa} for a recent investigation on the fate of matter coupling under disformal transformations)\footnote{Disformal transformations also allow one to consider black hole singularity from a different perspective, as recently discussed in \cite{Domenech:2019syf}.}.
Therefore, this mehtod allows one to construct new exact solutions which encode interesting deviations from GR. Consequently, the resulting solutions provide an interesting arena to investigate modification of the shadow of rotating compact objects induced by modified theories of gravity. Recent investigations in this direction have been reported in \cite{Khodadi:2020jij}. 

In this work, we shall extend the investigations initiated in \cite{BenAchour:2019fdf} to the axisymmetric framework. Concretely,  we will apply the solution-generating method here to produce a new rotating solution in DHOST theories. Transforming the stealth Kerr  ``seed'' solution by a disformal transformation \eqref{Intro:disformal} where the functions $A$ and $B$ are
constants, we present a new non-stealth rotating analytic solution, given by Eq.~(\ref{newsol}) below, which depend on three parameters: the mass $M$, the spin parameter $a$ and the disformal parameter $\alpha$ which encodes deviations form the Kerr metric. 
This new geometry is different from Kerr (it is no more Ricci flat for instance) but has similarities: it is asymptotically flat, it admits the two same Killing vectors  as the ones in Kerr, it has ergoregions very similar to those of Kerr, and it comes with the same  time-dependent scalar field (with constant kinetic term) as in the stealth Kerr solution. 
We explore geometrical properties of this family of novel rotating solutions: we argue that there is no new singularity compared to the usual Kerr black hole, we discuss the existence of event horizons computing  the disformed null vectors, and finally we study some
properties of its geodesics. 

The paper is organized as follows. In the following \S\ref{sec2}, we review basic results on stealth solutions in DHOST theories: we
give the  conditions for such solutions to exist and we quickly present the stealth Kerr solution. In \S\ref{sec3},
we construct and study solutions obtained from a disformal transformation \eqref{Intro:disformal}, where $A=A_0$ (without loss of generality we can fix $A_0=1$) and $B=B_0$ 
are constant, of this  stealth Kerr solution.
We start with reviewing useful results on disformal transformations which are the main
tool of our method. 
Then, we compute explicitly the novel solution and discuss some of its geometrical properties. 
In \S\ref{sec4} we conclude with a summary of the results and a discussion on open issues. 
In addition, to illustrate again the potentiality of the disformal solution-generating method, we  present and discuss in Appendix~\ref{AppB} another axisymmetric solution for DHOST theories obtained from a disformal transformation of the generalized Kerr solution of Einstein-Scalar gravity.

\medskip

\underline{Note added.} 
As we were writing this article, we learned that a similar (but complementary) article \cite{Anson}
was about to be posted on the arXiv. 
In \cite{Anson}, the authors study in more details the geometry of the disformed Kerr solutions and propose candidates for the event horizons. 
In the present paper, on the other hand, we clarify a class of DHOST theories in which the disformed Kerr solutions are exact solutions.
Our results and their results agree where we overlap.

\section{Stealth solutions in DHOST theories}
\label{sec2}
This section is devoted to review important results on stealth solutions in DHOST theories. We start, in \S\ref{sec2A}, with giving
the conditions that a DHOST theory must satisfy to have a stealth solution whose metric is also a solution of GR. In \S\ref{sec2B}, 
we describe more specifically the stealth Kerr solution which will be the ``seed''  to construct new rotating solutions in DHOST
theories in the subsequent section.

\subsection{Conditions of existence}
\label{sec2A}
The most general theory of quadratic DHOST theory \cite{Langlois:2015cwa} is described by the action 
\begin{equation}
\label{DHOST}
S=\int d^4x\sqrt{-g}\left(P(X,\phi)+Q(X,\phi)\, \Box \phi+F(X,\phi)\,R+\sum_{i=1}^{5}A_{i}(X,\phi)\, L_{i}\right)
\end{equation}
where the functions $A_{i},\,F,\,Q$ and $P$ depend on the scalar
field $\phi$ and its kinetic term $X\equiv \phi_\mu \phi^\mu$ with $\phi_\mu \equiv \nabla_{\mu}\phi$,
and $R$ is  the Ricci scalar.
The five elementary Lagrangians $L_{i}$ quadratic in second derivatives of $\phi$  are defined by 
\begin{eqnarray}
&&L_1 \equiv \phi_{\mu\nu} \phi^{\mu\nu} \, , \quad
L_2 \equiv (\Box \phi)^2 \, , \quad
L_3 \equiv \phi^\mu \phi_{\mu\nu} \phi^\nu\Box \phi \, , \quad \nonumber \\
&&L_4 \equiv  \phi^\mu  \phi_{\mu\nu} \phi^{\nu\rho} \phi_\rho \, , \quad
L_5 \equiv (\phi^\mu \phi_{\mu\nu} \phi^\nu)^2 \, ,
\end{eqnarray}
where we are using the standard notations $\phi_\mu \equiv \nabla_\mu \phi$ and $\phi_{\mu\nu} \equiv \nabla_\nu \nabla_\mu \phi$ for the first and second (covariant) derivatives of $\phi$. 
For the theory to propagate only one extra scalar degree of freedom in addition to the usual tensor modes of gravity, the functions $F$ and $A_i$ have to satisfy the so-called degeneracy conditions \cite{Langlois:2015cwa,Achour:2016rkg} while $P$ and $Q$ are totally free.
The degeneracy conditions can be derived for general higher-derivative theories in a systematic way~\cite{Motohashi:2016ftl,Motohashi:2017eya,Motohashi:2018pxg}.

It has been shown 
in \cite{Achour:2016rkg} that these DHOST  theories  can be classified  into three classes which are 
stable under general disformal transformations, i.e.\ transformations of the metric of the form
\beq
\label{disf}
g_{\mu\nu} \longrightarrow \tilde g_{\mu\nu}= A(X, \phi) g_{\mu\nu}+ B(X,\phi) \phi_\mu \, \phi_\nu\,,
\eeq
where $A$ and $B$ are arbitrary functions with the conditions that the two metrics are not degenerate. Notice that, when the disformal
transformation is not invertible, the DHOST theory after the transformation falls in the class of mimetic theories of gravity \cite{Chamseddine:2013kea,Takahashi:2017pje,Langlois:2018jdg}. See also \cite{Sebastiani:2016ras, Casalino:2018wnc, Gorji:2019rlm} for further details on mimetic gravity.

 The theories belonging to the first class, named class Ia in \cite{Achour:2016rkg}, can be mapped into a Horndeski form by applying a disformal transformation. 
The other two classes are not physically viable (either gradient instabilities of cosmological perturbations develop or tensor modes have pathological behavior)
\cite{Langlois:2017mxy}
and will not be considered in the present work.
Theories in class Ia  are labelled by the three free functions $F,A_1$ and $A_3$ (in addition to $P$ and $Q$) and 
the three remaining functions are given by the relations
\cite{Langlois:2015cwa}
\begin{eqnarray}
A_{2} & = & -A_{1} \, ,\label{deg1}
\\
A_{4} & = & \frac{1}{8\left(F+XA_{2}\right)^2}\Bigl(A_{2}A_{3}\left(16X^{2}F_{X}-12XF\right)+4A_{2}^{2}\left(16XF_{X}+3F\right) \nonumber\\
 &  & +16A_{2}\left(4XF_{X}+3F\right)F_{X} +16XA_{2}^{3}+8A_{3}F\left(XF_{X}-F\right)-X^{2}A_{3}^{2}F+48FF_{X}^{2}\Bigr) \, \label{funA4},\\
A_{5} & = & \frac{1}{8\left(F+XA_{2}\right)^2}\Bigl(2A_{2}+XA_{3}-4F_{X}\Bigr)\Bigl(3XA_{2}A_{3}-4A_{2}F_{X}-2A_{2}^{2}+4A_{2}^{3}F\Bigr) \, ,
\label{deg3}
\end{eqnarray}
where  $F_X$ denotes the derivative of  $F(X,\phi)$ with respect to $X$. Similarly $F_\phi$ denotes the partial derivative of $F$ 
 with respect to $\phi$ and the same notations will be used for all other functions as well.
 The above relations (\ref{deg1}-\ref{deg3}) are a direct consequence of the three degenerate conditions that
guarantee 
 only one scalar degree of freedom is present
\cite{Langlois:2015cwa,Langlois:2015skt}.
In conclusion, this means that all the DHOST theories we study here are characterized by five free functions of $X$ and $\phi$, which are $P$, $Q$, $F$, $A_1$ and $A_3$. Notice that we have implicitly supposed the condition $F+XA_{2}\neq 0$. Theories where
$F+XA_{2}= 0$ belong to the sub-class Ib which is not physically relevant \cite{Achour:2016rkg}. Finally, coupling to  external fields (perfect fluids, scalar fields, vector fields, etc.) can be done  by adding to the DHOST action an action $S_m$ where the external degrees of freedom are minimally coupled to the metric $g_{\mu\nu}$ (which is assumed to be the physical one).

\medskip

The Euler-Lagrange equations of DHOST theories are very complicated. Even though only one scalar degree of freedom comes with the usual two tensorial degrees of freedom of GR, these equations are higher order and can be up to fourth order in $\phi$ and third order in
$g_{\mu\nu}$. It is only in the Horndeski frame (where the DHOST theory falls in the Horndeski class) where the equations of motion are second order, but the external fields are no more minimally coupled to the metric and, even in that case, the equations still have a  very complex structure compared to GR. 
While the reduction of the higher-order Euler-Lagrange equations to a system of second-order differential equation for the case of static spherically symmetric space-time was performed explicitly in \cite{Takahashi:2019oxz}, the process is more involved for more general space-time.
Hence, finding exact solutions in DHOST theory is far from being an easy task and one thus usually makes assumptions to simplify the problem. 

Here we consider the following assumptions. First, we impose shift-symmetry which means that the DHOST action \eqref{DHOST} is unchanged by the transformation $\phi \rightarrow \phi+c$ where $c$ is a constant, and thus all the functions entering in the definition of \eqref{DHOST} depend on $X$ only.  Second, one
assumes the solution is such that  $X=X_0$ is a constant which drastically simplifies the modified Einstein equations. 
And finally, one looks for  
stealth solutions where the metric $g_{\mu\nu}$ is also a solution of the vacuum Einstein equations with a cosmological constant $\Lambda$,
\bea
G_{\mu\nu}+\Lambda g_{\mu\nu} = 0\, ,
\eea
where $G_{\mu\nu} \equiv R_{\mu\nu} - R g_{\mu\nu}/2$ is the Einstein tensor. One can go further
and requires that a given DHOST theory admits all GR solutions, and not only some of them, as the metric part of stealth solutions. This is the case if the following conditions hold \cite{Takahashi:2020hso},
\bea
\label{condStealth}
P + 2 \Lambda F = 0 \, , \quad
P_X + \Lambda (4 F_X - X_0 A_{1X}) = 0 \, , \quad
Q_X=0 \, , \quad
A_1 = 0 \, \quad
A_3 + 2 A_{1X} = 0 \, ,
\eea
where all these functions are evaluated on the solution $X=X_0$.
These conditions have been recently generalized to non-shift symmetric theories and to the case where matter is coupled to gravity minimally
\cite{Takahashi:2020hso}. 

Notice that these conditions are very strong and drastically restrict the set of DHOST theories. 
It was also argued that some stealth solutions lead to a problem of strong coupling \cite{Minamitsuji:2018vuw,deRham:2019gha} at the level of linear perturbations. 
Further, using the effective field theory framework it was shown in \cite{Motohashi:2019ymr} that perturbations about stealth solutions are strongly coupled, for de Sitter background in the decoupling limit, and for the Minkowski background even away from the decoupling limit, so long as we require evolution equation of perturbations to be second order.
Thus, in general the strong coupling is inevitable for asymptotically de Sitter or flat stealth solutions.  
Moreover, even if spacetime is different from de Sitter or Minkowski on superhorizon scales, the strong coupling is inevitable on subhorizon scales where the spacetime is nearly flat and hence the analysis of \cite{Motohashi:2019ymr} applies.
However, we can introduce a controlled detuning of the degeneracy condition, dubbed the scordatura mechanism in \cite{Motohashi:2019ymr}, to render the perturbations weakly coupled all the way up to a sufficiently high scale, as in the ghost condensate~\cite{ArkaniHamed:2003uy}. 
The Ostrogradsky ghosts associated with the scordatura is adjusted to show up only above the cutoff scale of the effective field theory.
It is also important to note that the scordatura does not change the properties of the stealth solutions of degenerate theories at astrophysical scales (similarly to the stealth solution \cite{Mukohyama:2005rw} in the ghost condensate). 
Thus, below we focus on stealth solutions in degenerate theories.

\subsection{The stealth Kerr solution in DHOST theories}
\label{sec2B}
The stealth Kerr solution in DHOST theories we shall use in this work as a seed was obtained in \cite{Charmousis:2019vnf}.  
It was derived for theories within the class Ia with no cubic galileon term where gravitational waves propagate at the speed of light ($c_{\rm GW}=c$), i.e.\
\bea
\label{cequal1}
A_1=A_2=0 \, , \qquad Q=0 \, .
\eea
Here, the condition $c_{\rm GW}=c$ implies $A_1=0$~\cite{Langlois:2017dyl}, and we have used the degeneracy condition $A_2=-A_1$ in \eqref{deg1}. 
Therefore, the stealth conditions \eqref{condStealth} simplify and become,
\bea
P(X_0) + 2 \Lambda F(X_0) = 0 \, , \quad
P_X(X_0) + 4 \Lambda F_X(X_0)  = 0 \, , \quad
A_3(X_0) = 0 \, .
\eea
The metric is the usual  Kerr solution of GR, or the de Sitter (dS)/Anti-de Sitter (AdS) Kerr solution 
in the presence of a non-zero cosmological constant $\Lambda=3/\ell^2$ ($\ell^2 >0 $ for dS or $\ell^2<0$ for AdS). In Boyer-Lindquist
coordinates $(t,r,\theta,\psi)$, it reads
\begin{align}
ds^2 = - \frac{\Delta_r}{\Xi^2 \rho} \left( dt - a \sin^2{\theta} \, d\psi \right)^2 + \rho \left( \frac{dr^2}{\Delta_r} + \frac{d\theta^2}{\Delta_{\theta}}\right) + \frac{\Delta_{\theta} \sin^2{\theta}}{\Xi^2 \rho} \left( a dt - \left(r^2 + a^2 \right) d\psi \right)^2
\label{Kerrmetric}
\end{align}
where $\Xi \equiv 1 + {a^2}/{\ell^2}$ is a constant, and the different functions entering in the metric are defined by
\begin{align}
\Delta_r & = \left( 1 - \frac{r^2}{\ell^2}\right) \left( r^2 + a^2\right) - 2 Mr \;, \qquad  \Delta_{\theta}  = 1 + \frac{a^2}{\ell^2} \cos^2{\theta} \;,\quad \rho  = r^2 + a^2 \cos^2{\theta} \, ,
\end{align}
while $M$ is the mass of the black hole and $a$ the angular momentum parameter satisfying the condition $a \leq M$.

The stealth Kerr black hole comes with a scalar hair whose highly non-trivial profile \cite{Charmousis:2019vnf} coincides with the
Hamilton-Jacobi potential associated to the Kerr geodesic equation and is given by,
\bea
\label{profilekerr}
\phi(t,r,\theta) = -Et + S_r(r) + S_{\theta} (\theta) \, , \qquad
S_r \equiv  \pm \int dr \, \frac{\sqrt{\cR}}{\Delta_r} \;, \quad S_{\theta} \equiv \pm \int d\theta \, \frac{\sqrt{\Theta}}{\Delta_{\theta}} 
\eea
where $E$ is a constant while the two functions $S_r$ and $S_\theta$ are defined, up to a sign ambiguity, as integrals involving the
radial and angular functions
\be
\label{func}
\cR(r) \equiv m^2 \left( r^2 + a^2\right) \left[ \eta^2 \left( r^2 + a^2\right) - \Delta_r\right]\;, \qquad \Theta(\theta) \equiv a^2 m^2 \sin^2{\theta} \left( \Delta_{\theta} - \eta^2 \right) \, ,
\ee
with $X_0=-m^2$ and $\eta \equiv \Xi E /m$. In fact, there are four different branches for the scalar field $\phi$ because of the freedom to choose the signs
of $S_r$ and $S_\theta$. It has been shown in \cite{Charmousis:2019vnf}  that one can make use of these branches to construct a scalar field solution which is regular and finite in an untrapped region as well as a trapped region (either a black hole region or a white hole region but not both), and in particular on the the black hole horizon (as well as on the cosmological horizons when $\ell^2 > 0$).  

In the particular case where there is no cosmological constant $\ell \rightarrow \pm \infty$, we have $\eta = 1$, then $\Theta$ vanishes, and
finally the scalar field does not depend on the variable $\theta$ anymore. As we are going to see in the following section, this is the case we 
will focus on when we consider disformal Kerr solutions to avoid several issues. Furthermore, the radial function $\cR(r)$ simplifies as well and becomes
\be
\label{funcR0}
\cR(r) \equiv 2 M m^2  r \left( r^2 + a^2\right)\; ,
\ee
with the condition $E=m$ which identifies the kinetic energy $X_0=-m^2$ to $E^2$ .

\section{Rotating solutions   beyond the stealth sector}
\label{sec3}
In this section, we will construct the new rotating solution in DHOST theories obtained from a disformal transformation
of the stealth Kerr black hole we have just described above. 
We will start, in \S\ref{sec3A}, by reviewing useful and general properties of disformal
transformations \eqref{Intro:disformal} on DHOST theories. Then, we will concentrate on such transformations that $A$ and $B$
are constant, and we will show how the stealth conditions \eqref{condStealth} transform under these ``constant'' disformal transformations giving the conditions for any DHOST theory to have the disformed Kerr black hole as a solution.
Finally, in \S\ref{sec3B} we will transform the stealth Kerr black hole and in \S\ref{sec3C} we study some geometrical properties of the disformed geometry, which is no more stealth.

\subsection{Constant disformal transformations and stealth conditions}
\label{sec3A}
When the metric $g_{\mu\nu}$ comes with a scalar field 
$\phi$, one can define the ``disformed'' metric $\tilde{g}_{\mu\nu}$ by~\cite{Bekenstein:1992pj}
\bea
\tilde{g}_{\mu\nu} = A(\phi,X) g_{\mu\nu} + B(\phi,X) \phi_\mu \phi_\nu \, ,
\eea
where $A$ and $B$ are arbitrary functions. One can show that the transformation is invertible when the condition $\partial_X(A/X+ B  ) \neq 0$
(and $A \neq 0$) is satisfied. 
We will always consider invertible disformal transformations here. Furthermore, as we restrict our study to shift-invariant DHOST theories, the functions $A$ and $B$ are supposed to depend on $X$ only.
Anticipating on the next section, we point that these two assumptions of i) invertibility and ii) shift symmetry lead to drastic simplifications when one considers the disformal mapping of seed solutions with a constant kinetic term $\tilde{X}=\tilde{X_0}$. Indeed, this automatically imposes that the disformal potentials $A$ and $B$ are constant 
\cite{BenAchour:2019fdf}. 

Disformal transformations on the metric induce transformations on DHOST actions. Given an action $\tilde S[\tilde{g}_{\mu\nu},\phi]$,
one defines a new action $S[g_{\mu\nu},\phi]$ by the identification,
\bea
S[g_{\mu\nu},\phi] = \tilde S[A(X) g_{\mu\nu} + B(X) \phi_\mu \phi_\nu,\phi] \, .
\eea
Interestingly, DHOST theories are stable under disformal transformations \cite{Achour:2016rkg} and the transformation
rules between the functions  (of $\tilde{X} \equiv \tilde{g}^{\mu\nu} \phi_\mu \phi_\nu$) $\tilde{P}$, $\tilde{Q}$,
$\tilde{F}$ and $\tilde{A}_I$ entering in the definition of the action $\tilde{S}[\tilde{g}_{\mu\nu},\phi]$ on one side, 
and  the functions (of $X = g^{\mu\nu} \phi_\mu \phi_\nu$) $P$, $Q$, $F$ and $A_I$ defining $S[g_{\mu\nu},\phi]$  on the other side
are given in \cite{Achour:2016rkg}. 

As it turns out, these rules, which are rather complicated, simplify drastically when one considers 
constant disformal transformations where $A=A_0$ and $B=B_0$ do not depend on $X$ anymore. As pointed above, this is the case when considering invertible and shift symmetric disformal mapping of seed solution with constant kinetic term. After a straightforward
calculation, one shows that the k-essence, the cubic galileon and the Ricci terms transform as follows,
\bea
P = \tilde{P} \, , \qquad
Q =  A_0  \int dX \, N {\tilde Q}_X \, , \qquad F  =\frac{A_0}{N}\tilde{F}  \, ,
\eea
while the functions $A_I$ entering in the quadratic part of the Lagrangian transform as,
\bea
&&A_1 = N (B_{0} \tilde{F} + N^2 \tilde{A}_1) \, , \quad
A_2  = N(- B_{0} \tilde{F} + N^2 \tilde{A}_2) \, ,\nonumber \\
&& A_3  = \frac{N}{A_0} \left[  - 4 A_0 B_0 \tilde{F}_X - 2 B_0 N^4\tilde{A}_2 +N^6 \tilde{A}_3\right] \, , \nonumber  \\
&&A_4  = \frac{N}{A_0} \left[- N^2 B_0^2 \tilde{F} + 4 A_0 B_0 \tilde{F}_X - 2 N^4 B_0 \tilde{A}_1+ {N^6} \tilde{A}_4 \right] \, , \nonumber  \\
&&A_5  = \frac{N^7}{A_0^2}\left[ B^2_{0} ( \tilde{A}_1 + \tilde{A}_2) + N^2 B_{0}( \tilde{A}_3 - \tilde{A}_4) +N^4 \tilde{A_5} \right] \, ,
\eea
where we introduced the factor
\bea
N \equiv {A_0}^{1/2}{(A_0 + X B_0)}^{-1/2} \, .
\eea
We recall that ``tilde'' functions $\tilde{P}$, $\tilde{Q}$, $\tilde{F}$, $\tilde{A}_I$, in the right-hand side of the previous equations are viewed as functions of $X$ via the relation,
\bea
\label{X}
\tilde{X} = \frac{X}{ A_0 + X B_0} \, .
\eea

\medskip

Now, we assume that the theory $\tilde{S}[\tilde{g}_{\mu\nu},\phi]$ satisfies the conditions \eqref{condStealth} to have a stealth 
solution where $\tilde{X}_0$ is constant and  $\tilde{g}_{\mu\nu}$ is the Kerr metric recalled above \eqref{Kerrmetric}. Then, we 
want to translate these conditions in terms of the functions entering into the action $S[g_{\mu\nu},\phi]$. 
We first remark that under constant disformal transformation $X_0$ is also constant when $\tilde{X}_0$ is constant.
After a direct calculation, from \eqref{condStealth} we obtain
\bea
&&P + \frac{2 \Lambda N}{A_0} F = 0 \, , \qquad
{\partial_X} \left[ P + \frac{\Lambda}{A_0} \left( 4 + \frac{B_0 X_0}{N}\right) F - \frac{\Lambda X_0}{N^3} A_1\right] = 0  \, , \label{cond1}\\
&& Q_X = 0 \, , \qquad A_1 - \frac{N^2 B_0}{A_0} F = 0 \, , \label{cond2}
\eea
together with the remaining more complicated condition 
\bea
\frac{A_0}{2N} A_3 +  \left( 2 B_0 N_X + \frac{B_0 N^2}{A_0}\right)F + 2 B_0 N F_X + B_0 N A_2 + \frac{N^8}{A_0}
\partial_X \left( \frac{A_1}{N^3} - \frac{B_0}{A_0} \frac{F}{N}\right)=0 \, , \label{cond3}
\eea
which comes from the last equation of \eqref{condStealth}. Let us recall that these equations holds only when they  are
evaluated on $X_0$. As a consequence, any DHOST theories which satisfy all these conditions admit disformal stealth solutions, which
are in general non-stealth. Obviously, these conditions reduce to \eqref{condStealth} for a trivial disformal transformation where
$A_0=1$ and $B_0=0$.

In the special case where the theory $\tilde{S}$ satisfies, in addition, the conditions \eqref{cequal1} which insure that gravitational waves
propagate at the speed of light, the disformed theory $S$ satisfies in turn, 
\bea
A_1= - A_2 = N^2 \frac{B_0}{A_0} F \, , \qquad Q=0 \, .
\eea
Hence, \eqref{cond2} are automatically satisfied and the first equation in \eqref{cond1} is unchanged. 
The last  conditions in  \eqref{cond1}  and \eqref{cond3} are also simplified according to
\bea
\partial_X \left( P + \frac{4 \Lambda}{A_0} F\right)=0 \, , \qquad
\frac{A_0}{2 N B_0} A_3 + \left(2N_X + \frac{N^2}{A_0}(1-NB_0) \right) F + 2N F_X =0 \, .
\eea 
In that case, the theory $S$ admits the disformed Kerr black hole  we are going to describe now as a solution.

\subsection{Disformal Kerr solution: construction and preliminary properties}
\label{sec3B}

Considering the previous stealth Kerr-(A)dS seed solution with a constant kinetic term, we turn now to generate a new non-stealth solution whose metric takes the form
\bea
\label{disKerrformal}
g_{\mu\nu} \; = \; \tilde{g}_{\mu\nu} - B_0 \, \phi_\mu \phi_\nu \, ,
\eea
where $\tilde{g}_{\mu\nu} $ is the Kerr metric \eqref{Kerrmetric}. Without loss of generality, we have fixed $A_0=1$, otherwise the metric would simply get a global physically irrelevant constant conformal factor. If the scalar field $\phi$ depends on the angular variable $\theta$, then 
the  disformed Kerr metric \eqref{disKerrformal} acquires new components, among which
\bea
g_{t \theta} =  \pm B_0 E \frac{\sqrt{\Theta}}{\Delta_\theta} \, , 
\eea
where the expression of $\Theta(\theta)$ and $\Delta_\theta(\theta)$ has been recalled  in \eqref{func}.
Such a  term depends on the radial variable $r$ and then they do not vanish at infinity. As a consequence, one cannot expect
that the disformed metric is asymptotically flat, dS or AdS. To avoid this pathological behavior, we require that the scalar field does not
depend on $\theta$ which implies necessarily the vanishing of the cosmological constant $\ell \rightarrow \pm \infty$, then $\eta=1$ and $E=m$ (as
a consequence of $\Theta=0$). Hence, from now on, we consider only this case which, as recalled before \eqref{funcR0},  corresponds to scalar field of the form
\bea
\label{profilekerr}
\phi(t,r) = -m t + S_r(r)  \, , \qquad
S_r =  \pm \int dr \, \frac{\sqrt{\cR}}{\Delta} \, , \qquad
\Delta = r^2+a^2-2Mr \; , 
\eea
where, for simplicity, we have omitted the subscript $r$ in $\Delta$ (as there is no more possible ambiguity).
Thus, the disformal transformation (with $A_{0} =1$) leads to the new solution 
\bea
\label{newsol}
ds^2 &=&- \frac{\Delta}{\rho} \left( dt - a \sin^2{\theta} \, d\psi \right)^2 + \frac{\rho}{\Delta} {dr^2} + \rho \, {d\theta^2} + \frac{\sin^2{\theta}}{\rho} \left(a \, dt - \left(r^2 + a^2 \right) d\psi \right)^2 \nonumber \\
&& + \alpha \left( dt  \pm  {\sqrt{2Mr(r^2+a^2)}}/{\Delta} \, dr\right)^2 \, ,
\eea
with $\alpha \equiv -B_0 m^2$ while the scalar field profile remains unchanged (\ref{profilekerr}). 
The inverse disformed Kerr metric is given by
\bea
\label{inversedisf}
g^{\mu\nu} \, = \, \tilde{g}^{\mu\nu} + \frac{\alpha}{m^2(1-\alpha)} \phi^\mu \phi^\nu \, ,
\eea
where $\tilde{g}^{\mu\nu}$ is the inverse Kerr metric while the only non-vanishing components of $\phi^\mu = \tilde{g}^{\mu\nu} \phi_\nu$ 
are,
\bea
\phi^t =  \frac{m}{\Delta}\left( r^2+a^2 + \frac{2M r a^2 \sin^2\theta}{\rho}\right) \, , \quad
\phi^r =   m \frac{\sqrt{2M r (r^2+a^2)}}{\rho}\,, \quad
\phi^\varphi =m\frac{2aMr }{\Delta \rho} \, .
\eea
Therefore, the disformal transformation of the stealth Kerr solution provides a new \textit{non-stealth} exact solution which is parametrized, in addition to the  mass and angular momentum parameters $(M, a)$ of the Kerr family, by one new deformation parameter $ \alpha$ which encodes precisely the deviations from GR. The apparent $\pm$ ambiguity in \eqref{newsol} can be absorbed thanks to simple redefinitions of 
$t$ and $a$ which are replaced by $\pm t$ and $\pm a$. Hence, we can safely fix the sign to $+$ from now on without loss of generality.

\medskip

At infinity where $r\rightarrow + \infty$, the disformed Kerr metric becomes equivalent to,
\bea
\label{asymptomet}
ds^2 &\simeq& -\left( 1-\frac{2M_1}{r}\right)dt^2 + \left( 1-\frac{2M_2}{r}\right)^{-1} dr^2 + r^2 (d\theta^2 + \sin^2\theta \, d\varphi^2) \nonumber \\
&+& 2\alpha \sqrt{\frac{2M_1}{r}} dr\,dt + {\cal O}\left( \frac{1}{r^2}\right) \, , 
\eea
where we introduced the notations,
\bea
M_1 \equiv \frac{M}{1-\alpha} \, , \quad M_2\equiv(1+\alpha)M \, ,
\eea
and we rescaled the time coordinate $t$ by $\sqrt{1-\alpha}$. Note that the coefficient $\alpha$ modifies the black hole mass in the matrix elements $g_{tt}$ and $g_{rr}$ in a different way in Schwarzschild coordinates as $M_1 \neq M_2$. These masses agree at the first order in the parameter $\alpha$. Moreover, while the cross term $g_{tr}dt dr$ induced by the disformal transformations decays in the asymptotic regime, one can show that it cannot be removed by a coordinate change without introducing new off-diagonal terms, such that the new solution is not circular. This property appears a the key novelty of this new exact solution. See \cite{Anson} for more details on this point. Hence, the metric is asymptotically flat but, contrary to the Kerr metric, the disformed one is not equivalent to the Schwarzschild metric at infinity
essentially because the difference between the masses $M_1$ and $M_2$. Nevertheless, the deviations introduced by the presence of the cross term proportional to $dr dt$ in the metric become manifest only at next-to-leading order in the asymptotic expansion \cite{Anson}.

\subsection{Some  properties of the disformed Kerr space-time}
\label{sec3C}
In this section, we quickly discuss geometrical properties of the disformed Kerr space-time. First of all,  we say a few words on its singularities. Even though the metric is 
singular in Boyer-Lindquist coordinates when $\Delta=0$ (at the values $r_\pm = M \pm \sqrt{M^2-a^2}$ of the radial coordinate), this is 
not a physical singularity but only a coordinate singularity exactly as in the usual Kerr black hole. Indeed, this can be seen immediately from the expressions of the curvature invariants,
\begin{align}
R & =  \frac{ \alpha }{1-\alpha} \frac{6 a^2 M r}{\rho^3} \left(\cos^2{\theta} - \frac{1}{3} \right)\, , \\
\label{CI2}
R_{\mu\nu} R^{\mu\nu} & = - \frac{\alpha^2}{\left( 1-\alpha \right)^2} \frac{18 a^4 M^2 }{\left( r^2 + a^2\right)\rho^6} \; P_1(r, \theta, a) \, ,\\
\label{CI3}
R_{\mu\nu\rho\sigma} R^{\mu\nu\rho\sigma} & = \frac{48 M^2}{ \left( 1-\alpha \right)^2 \left( r^2 + a^2\right) \rho^6} \; P_2 \left( r, \theta,a\right) \, ,
\end{align}
where  the functions $P_1(r, \theta, a)$ and $P_1(r, \theta, a)$ are given in the Appendix~\ref{AppA}.
As for the Kerr metric, the disformed Kerr geometry is singular at $\rho=0$ only. Nonetheless, one important difference between the disformed and the usual Kerr metric is that the disformed geometry ($\alpha \neq 0$) is, interestingly, no longer Ricci flat.

Then, we see immediately that the disformed geometry admits the two same Killing vectors $\xi_t \equiv \partial_t$ and $\xi_\varphi \equiv \partial_\varphi$ as in the Kerr black hole because none of the
coefficients of the metric depend on $t$ and $\varphi$. As a consequence, we can look at the positions of the ergospheres, i.e.\ the hypersurfaces where the Killing vector field $\xi_t$ is null, i.e.\
\bea
\xi_t \cdot \xi_t = g_{tt} = 0  \qquad \Longleftrightarrow \qquad r^2 -2 {M_1} r + a^2 \cos^2\theta = 0 \, ,
\eea
where $M_1=M/(1-\alpha)$ as above \eqref{asymptomet}.
As a consequence, the disformed Kerr metric admits, as the usual Kerr metric, an outer and an inner ergospheres denoted 
respectively by ${\cal E}^+$ and ${\cal E}^-$ whose positions are given by the same formulae as the Kerr ones,
\bea
r=r_{{\cal E}^\pm}(\theta)= M_1 \pm \sqrt{M_1^2 - a^2 \cos^2\theta} \, ,
\eea
with the difference that the mass of the black hole has now been rescaled.  The ergoregions are defined similarly and one expects the possibility for a Penrose process (with an energy extraction mechanism) to exist in this geometry as well.

Now, let us consider the null directions. Indeed, computing the null directions is particularly interesting  to understand the causal structure of a metric and to see whether a metric $g_{\mu\nu}$ describes a black hole (or more generally possesses  horizons). These vectors enable us, in particular, to compute  light rays (the principal null geodesics) in the space-time and also to characterize the properties of horizons. The normalized (future directed) principal null vectors are denoted by
$\ell_\pm^\mu$ and satisfy the normalization conditions,
\bea
g_{\mu\nu}  \, \ell_\pm ^\mu \, \ell_\pm^\nu \; = \; 0 \; , \qquad g_{\mu\nu}  \, \ell_+^\mu \,  \ell_-^\nu \; = \; - 1 \, .
\eea
These conditions do not define completely (and then uniquely) the null vectors which can be rescaled according to $\ell_\pm \rightarrow \mathcal{N}^{\pm 1} \ell_\pm$ where $\mathcal{N}$ is an arbitrary (non-vanishing) function. 

In the case of the Kerr metric $\tilde{g}_{\mu\nu}$, the null vectors are well-known and are given by,
\bea
\tilde{\ell}_+^\mu \partial_\mu \equiv \frac{r^2+a^2}{\Delta} \partial_t + \partial_r + \frac{a}{\Delta} \partial_\varphi \,, \qquad
\tilde{\ell}_-^\mu \partial_\mu \equiv  \frac{r^2+a^2}{2 \rho} \partial_t - \frac{\Delta}{2\rho}\partial_r + \frac{a}{2 \rho} \partial_\varphi \, .
\eea
Interestingly, we see that, at the horizons where $\Delta=0$, these two null vector fields are proportional to the Killing vector
\bea
\frac{\Delta}{r^2+a^2} \tilde{\ell}_+ = \frac{2\rho}{r^2+a^2} \tilde{\ell}_- =\partial_t + \Omega_H \, \partial_\varphi \, , 
\eea
where $\Omega_H=a/(2M r_\pm)$ is a constant whose value depends whether we are considering the outer ($r=r_+$) or the inner ($r=r_-$)
horizon. In fact, requiring that a Killing vector is null characterizes completely the horizons in the Kerr geometry where the event horizons are thus also Killing horizons. 

One can easily see that the principal directions $\ell_\pm$ of the disformed metric $g_{\mu\nu}$ are also ``disformed'' in the sense
that they are now given by,
\bea
\ell_\pm^\mu = \tilde{\ell}_\pm^\mu + \beta  (\phi_\alpha    \tilde{\ell}_\pm^\alpha) \, \phi^\mu \, , \qquad \beta \equiv \frac{(1- B_0  X)^{-1/2} -1}{X} =  \frac{1-(1- \alpha)^{-1/2} }{m^2} \, ,
\eea
where $\phi^\mu = \tilde{g}^{\mu\nu} \phi_\nu$ and $X=-m^2$ here. Notice that these formulae generalize immediately to any disformal transformation (when $X$ and $B_0$ are not necessarily constant) of  an arbitrary metric $g_{\mu\nu}$.  
Interestingly the effect of the disformal transformation on the null directions is a shift of the usual Kerr null vectors in the direction of the gradient of the scalar field $\phi^\mu$. Everything happens as if the scalar field is somehow drifting the light rays.

The explicit expressions of the null directions of the 
disformed Kerr metric can be easily written from the relations
\bea
\phi_\alpha \tilde \ell_+^\alpha & = & -\frac{m \sqrt{r^2+a^2}}{\sqrt{r^2+a^2} + \sqrt{2Mr}} \, , \qquad
\phi_\alpha \tilde \ell_-^\alpha =- m\frac{r^2+a^2 + \sqrt{2Mr(r^2+a^2)}}{2 \rho} \, ,
\eea
which enable us to obtain, after a direct calculation, 
\bea
\ell_+ & = & \left[ \frac{r^2+a^2}{\Delta} + \beta m (\phi_\alpha \tilde \ell_+^\alpha) \left(1+ \frac{\cR}{m^2 \rho \Delta} \right)\right] \partial_t \nonumber \\
&+& \left[ 1 + \beta (\phi_\alpha \tilde \ell_+^\alpha) \frac{\sqrt{\cR}}{\rho}\right] \partial_r  - \frac{a}{\Delta}\left[1 + \beta m (\phi_\alpha \tilde \ell_+^\alpha) \frac{2  M r}{\rho}  \right] \partial_\varphi \, ,   \\
\ell_- & = & \left[ \frac{r^2+a^2}{2\rho} + \beta m (\phi_\alpha \tilde \ell_-^\alpha) \left(1+ \frac{\cR}{m^2 \rho \Delta} \right)\right] \partial_t
\nonumber \\
&+ & \left[ -\frac{\Delta}{2 \rho} + \beta (\phi_\alpha \tilde \ell_-^\alpha) \frac{\sqrt{\cR}}{\rho}\right] \partial_r - \frac{a}{2 \rho}\left[1 + \beta m (\phi_\alpha \tilde \ell_-^\alpha) \frac{2  M r}{\Delta}  \right] \partial_\varphi   .
\eea
If we proceed as in the Kerr case, we would look at the regions where $\ell_+$ or $\ell_-$ becomes proportional to Killing vectors. For 
$\ell_+$, we obtain the condition,
\bea
\Delta \left[ 1 + \beta (\phi_\alpha \tilde \ell_+^\alpha) \frac{\sqrt{\cR}}{\rho}\right] \; = \; 0 \, ,
\eea
which fixes $r$. Interestingly, $r_\pm$ are solutions but there are extra non-trivial solutions which are given by $r=F(\theta)$, i.e. $r$
is a function of $\theta$ and whose limit $\alpha \rightarrow 0$ is not defined. 
For $\ell_-$, we obtain the condition,
\bea
  -{\Delta} + 2 \beta (\phi_\alpha \tilde \ell_-^\alpha) \sqrt{\cR} \; = \; 0 \, ,
\eea
which also fixes $r$ at some non trivial function of $\theta$. In both cases, both null vectors reduce to a vector field proportional to,
\bea
\partial_t + \Omega_\pm \, \partial_\varphi \, ,
\eea
where $\Omega_\pm$ is no more a constant and depends on $\theta$. 
Therefore, none of the principal null directions reduce to  Killing vectors in some hypersurfaces, and hence the horizons 
(if they exist) cannot be Killing vectors\footnote{This has also been observed in another but equivalent way in \cite{Anson}}. This is an important difference with the Kerr geometry. 
Furthermore, one can easily check that the hypersurfaces of constant $r$ are not null because the norm of their normal vector $\partial_r$ is given by \eqref{inversedisf}
\bea
g^{rr}= \frac{\Delta}{\rho} + \frac{\alpha}{1-\alpha}\frac{2Mr(r^2+a^2)}{\rho^2}  \, ,
\eea
and therefore depends on $\theta$ through $\rho=r^2+a^2 \cos^2\theta$. As a consequence, the horizons of the disformed Kerr metric cannot be obtained in the way we get the
Kerr horizons. 

The problem of finding event horizons seems complicated but it has been initiated  very recently in \cite{Anson} where first candidates  has been proposed and analyzed. The basic idea is rather simple and consists in looking at null hypersurfaces defined by an equation of the form $F(r,\theta)=0$  with a $\theta$-dependency contrary to the Kerr case. We assume that we can locally solve $r$ as a function of $\theta$ and restrict ourselves to separable functions of the form $F(r,\theta)=r+F(\theta)$.
 The condition that such an hypersurface is null implies that its normal vector  
$(0,1, \partial_\theta F,0)$ is also null (by definition), which leads to a non-linear differential equation for $F(\theta)$,
\bea
g^{rr}  + g^{\theta \theta} \left(\frac{dF}{d\theta}\right)^2 \, = \, 0 \,  \Longleftrightarrow \, \left[{\Delta} + \frac{\alpha}{1-\alpha}\frac{2Mr(r^2+a^2)}{\rho} \right] +  \left(\frac{dF}{d\theta}\right)^2 \, = \, 0 \, ,
\eea
where we used \eqref{inversedisf} for the coefficients of the inverse disformed metric and $r=-F(\theta)$ everywhere in this equation.  It is the 
same equation as Eq.(23) in  \cite{Anson}. The geometry of this null hypersurface is subtle as the detailed and very interesting analysis  in \cite{Anson} shows, but it is still an open issue to show whether it is an event horizon or not. Computing its expansions may help and we hope to study this issue in details in a future work. Nevertheless, it is worth emphasizing that the characterization of quasi-local horizon through the expansions of the null directions is slicing dependent, and the choice of the null directions and thus of the $2$-surface foliating our geometry is therefore ambiguous as different choices might allow to identify different quasi-local (not necessarily null) horizons. See \cite{Faraoni:2016xgy, Schnetter:2005ea} for detailed discussions on this point.

\medskip

We finish with a quick discussion on the geodesic equations in the disformed Kerr background. Following the same method as in the case
of the Kerr metric, the geodesic equations can be obtained from the Hamilton-Jacobi equation for the ``action'' $S$,
\bea
H(x^\mu, {\partial_\mu S}) + \frac{\partial S}{\partial \lambda} = 0 \, , \qquad H(x^\mu,p_\mu) \equiv \frac{1}{2} g^{\mu\nu} p_\mu p_\nu \, ,
\eea
where $\lambda$ is the affine parameter along the geodesic. Due to the invariance of the disformed metric (whose components do not depend neither on $t$ nor on $\varphi$), the action $S$ takes the form
\bea
S(t,r,\theta,\varphi) \; = \; \frac{1}{2} \mu^2 \lambda + p_t t + p_\varphi \varphi + \Phi(r,\theta) \, ,
\eea
where $\mu$, $p_t$ and $p_\varphi$ are the standard constants of motion. A straightforward calculation shows that $\Phi(r,\theta)$
satisfies the differential equation,
\bea
0&=&\left[ \mu^2 r^2 + \Phi_r^2 \Delta - \frac{(r^2+a^2)^2}{\Delta} - \frac{4Mra}{\Delta} p_t p_\varphi - \frac{a^2}{\Delta} p_\varphi^2 \right] \nonumber \\
&+&\left[ \mu^2 a^2 \cos^2\theta + \Phi_\theta^2 + a^2 p_t^2 \sin^2\theta + \frac{p_\varphi^2}{\sin^2\theta}\right] \label{HJeq} \\
&-&\frac{\alpha}{(1-\alpha m^2)\rho} \left[ \frac{m}{\Delta} (r^2+a^2)^2 p_t + \sqrt{\cR} \Phi_r - ma^2 p_t \sin^2\theta \right] ^2 \, ,
\nonumber
\eea
where $\Phi_r\equiv \partial \Phi/\partial r$ and $\Phi_\theta\equiv \partial \Phi/\partial \theta$.
In the case where $\alpha=0$, the equation is clearly separable as the first line depends only on $r$ while the second one depends on $\theta$ only. This makes the geodesic equation integrable. Furthermore, the separability of the Hamilton-Jacobi equation is intimately linked to the existence of the famous  Carter constant and of a hidden symmetry of the Kerr metric (associated to a Killing tensor)~\cite{Carter:1968rr}. 
When $\alpha \neq 0$, the Hamilton-Jacobi equation is no more separable in the Boyer-Lindquist coordinates and it is very likely that the geodesic equation is no more integrable and one cannot find a ``disformed'' Carter constant associated to the
disformed Kerr metric. Interestingly, there is an obvious solution of the equation \eqref{HJeq} given by 
\bea
\Phi= z \int dr \, \frac{\sqrt{\cR}}{\Delta} \, ,
\eea 
where $z$ is a
constant when the integration constants $\mu,p_t$ and $p_\varphi$ coincide with those of the scalar field according to,
\bea
p_t=-zm \, , \quad
p_\theta=0 \, , \quad
p_\varphi=0 \, , \quad
\mu^2 = \frac{z^2m^2}{1-\alpha m^2} \, .
\eea
In that case, the geodesic follows exactly the gradient of the scalar field. 

\section{Discussion and perspectives}
\label{sec4}
The disformal-generating method that was recently introduced in \cite{BenAchour:2019fdf} appears to be very useful to construct new exact
solutions in DHOST theories. It enables us, in this paper, to construct the first non-stealth rotating solution in DHOST theories  where the geometry is given by a disformed Kerr metric while the scalar field $\phi(t,r)$ has a non-trivial profile with a constant kinetic density $X$. Even
though this solution is equivalent to the stealth Kerr solution of \cite{Charmousis:2019vnf} in vacuum, it becomes physically inequivalent when one considers coupling to matter (for instance, the geodesic motion of test particles is different from the geodesic motion in the Kerr black hole as we briefly showed in the last part of the paper). 

We started with a quick review on quadratic DHOST theories and the conditions of existence of stealth solutions where the metric is
also a solution of GR. Then, we found the general conditions for a DHOST theory to have a ``disformed'' stealth solution 
where the metric is a ``disformed'' solution of GR while the scalar field has a non-trivial profile. This is an important result because 
it allows one to identify DHOST theories which admit a disformed stealth metric as a solution. Then, we performed a disformal transformation of the theories which admits a stealth Kerr solution and we obtained a family of scalar-tensor theories which admits disformed Kerr solutions. We have restricted ourselves to invertible and shift symmetry disformal transformation. As a consequence, working with a seed stealth Kerr solution with constant kinetic term imposes that $A=A_0$
and $B=B_0$ are constant, providing a drastic simplification. Under these assumptions, we have obtained the first non-stealth rotating solution in DHOST theories where the metric depends on 
three parameters which are the usual mass $m$ and spin parameter $a$ together with a new deformation parameter $\alpha$. 

We analyzed some geometrical properties of the new metric. It is easy to see that, contrary to the Kerr metric, the disformed Kerr metric is not Ricci flat but (in the case where the scalar field does not depend on $\theta$ in the Boyer-Lindquist coordinates system) it remains asymptotically flat, the metric still has two Killing vectors associated to the fact that the metric components are independent of
$t$ and $\varphi$ (still in Boyer-Lindquist coordinates),  there is the same ring singularity at $\rho=0$ as in Kerr, and the space-time
admits ergospheres and ergoregions very similar to those in Kerr space-time. However, there are  important differences with Kerr. First, 
we showed  that the null directions are disformed, which lead to modifications of the structure of the horizons. In particular, if horizons exist,
there can no longer be Killing horizons and cannot be given by $r=\rm const$ in Boyer-Lindquist coordinates. So far, we have no proof that the disformed solution is a black hole. These issues have been analyzed 
in the recent paper \cite{Anson} where a candidate for the event horizons has been proposed. However, understanding the geometry and more particularly the causal structure of the disformed Kerr metric deserve to be studied in details, what we hope to do next. 
Many other questions, that we hope to address in future works,  remain open: analyzing the geodesic motion, studying the thermodynamics, etc.

Finally, to illustrate again the potentiality of the disformal solution-generating method, we  present in addition another axisymmetric solution for DHOST theories obtained from a disformal transformation of the generalized Kerr solution of Einstein-Scalar gravity in the Appendix~\ref{AppB}.

\medskip


\acknowledgments
The work of JBA was supported by Japan Society for the Promotion of Science (JSPS) Grants-in-Aid for Scientific Research (KAKENHI) No.\ 17H02890. 
HM was supported by JSPS KAKENHI No.\ 18K13565.
The work of SM was supported by JSPS KAKENHI No.\ 17H02890, No.\ 17H06359, and by World Premier International Research Center Initiative, MEXT, Japan. 
KN acknowledges support from the CNRS grant 80PRIME  and  thanks the Laboratory of Physics at the ENS in Paris for hospitality during this exceptional period. 
We would like to thank the authors of \cite{Anson} for letting us know about their work prior to submission to arXiv and KN is especially grateful to Christos Charmousis and Eugeny Babichev for our discussions on the disformed Kerr geometry.

\appendix
\section{Explicit forms of $P_1(r,\theta, a)$ and $P_2(r,\theta, a)$}
\label{AppA}

The functions $P_1(r,\theta, a)$ and $P_2(r,\theta, a)$ entering in the formulas of the curvature invariants (\ref{CI2}) and (\ref{CI3}) are  given by,
\bea
P_1 (r,\theta, a) & =  & \left( z^2 -1\right) z^2 a^4 - \frac{4a^2 a^2}{9} \left( z^4 + \frac{5}{2} z^2 + \frac{5}{2}\right) - r^4 \left( z^4 + \frac{5}{9} z^2 + \frac{10}{9} \right) \, , \\
P_2 (r,\theta, a) & =  & \left[ 
\left( 
\left(\alpha -1 \right) z^4 - \frac{3z^2 \alpha^2}{2} + \frac{3\alpha^2}{2}\right) z^2 a^8  \right.  \nonumber \\
&&  \left. + 
\left( \left(  
\left( \alpha -1\right) z^6- \left( \frac{\alpha^2}{12} + 13 \alpha - 15\right) z^4  +  \left( \frac{13}{6} \alpha - 1\right) \alpha z^2 + \frac{19\alpha^2}{12} \right) r^2 a^6 \right) r^2 a^6  \right. \nonumber \\
&& \left. + \left( \left( \alpha^2 - \frac{52}{3} \alpha + 20\right) z^4 + \left( \frac{16}{9} \alpha^2 + 12 \alpha - 20\right) z^2 + \frac{19 \alpha^2}{9} + \frac{20 \alpha}{9}\right) \frac{3r^4 a^4}{4} \right. \nonumber  \\
&&   \left. 
+ \; 10  \left( \left( \alpha - \frac{3}{2}\right) z^2 + \frac{\alpha}{6} + \frac{1}{10}\right) a^2 r^6 + r^8 \right] \, ,
\eea
where we have used the notation $z=\cos{\theta}$ for simplicity.

\section{A rotating naked singularity solution in DHOST}
\label{AppB}

In this appendix, we construct  another exact solution of the DHOST theories starting from a seed describing the generalization of Kerr spacetime with a scalar source in General Relativity found in \cite{Bogush:2020lkp}. 

As expected from the no hair theorem, this exact solution of the Einstein-Scalar system does not describe a black hole but a rotating naked singularity. Nevertheless, it provides an interesting seed as it allows to work with a generalized Kerr geometry associated to a scalar source whose kinetic term is no longer constant. Moreover, the static limit of this  solution reduces to the well known quadrupolar metric and allows us to generate a DHOST generalization of this axisymmetric deformation of the Schwarzschild geometry.

\subsection{The rotating solution of Einstein-Scalar gravity}

Let us first describe the seed solution of the massless Einstein-Scalar system whose action reads
\be
\cS = \int \sqrt{|g|} \left( \frac{R}{16 \pi G} -  \frac{1}{2} g^{\mu\nu} \phi_{\mu} \phi_{\nu}  \right) \, .
\ee
The field equations are given by
\begin{align}
G_{\alpha\beta}  = 8\pi G \left(\frac{1}{2} \phi_{\mu} \phi^{\mu} \, g_{\alpha\beta}  - \phi_\alpha \phi_\beta\right)  \;, \qquad \Box \phi  = 0 \, .
\end{align}
As usual, the equation of the scalar field is a consequence of the Bianchi identity and the Einstein equation.
Recently, an exact stationary solution of these  equations was constructed in \cite{Bogush:2020lkp}. This solution was obtained by exploiting the well known hidden symmetries of vacuum axisymmetric solutions of GR \cite{Ehlers:1957zz, Matzner:1967zz, Geroch:1972yt, Hoenselaers:1985qk} as well as solution-generating method based on the Einstein-Maxwell system \cite{Galtsov:1995mb, Clement:1997tx}. The resulting geometry provides a generalization of the Kerr geometry with a scalar source. Explicitly, the metric  takes the form
\begin{align}
ds^2 = - f \left( dt - \omega d\psi \right)^2 +  \frac{h_{ij}}{f} dx^i dx^j  \, ,
\end{align}
where the functions $f, \omega$ and $\Delta $ entering in the metric coefficients are given by
\begin{align}
f \equiv 1 - \frac{2Mr}{\rho}\;, \qquad 
 \omega  \equiv - \frac{2a M r \sin^2{\theta}}{\Delta - a^2 \sin^2{\theta}}\;, \qquad  \Delta  \equiv \left(r- M\right)^2 - b^2 \, ,
\end{align}
with $b \equiv \sqrt{M^2 - a^2}$ and $\rho = r^2 + a^2 \cos^2{\theta}$. The remaining spatial part of the line element is explicitly given by,
\be
h_{ij} dx^i dx^j \equiv H \left( dr^2 + \Delta \, d\theta^2 \right) + \Delta \sin^2{\theta} \, d\psi^2 \, 
\ee
where the function $H$ is given by
\be
H \equiv 
\frac{\rho}{\Delta}f \zeta \;, \qquad \text{with} \qquad \zeta = \left( 1 + \frac{b^2}{\Delta} \sin^2{\theta}\right)^{- \Sigma^2 /b^2}.
 \ee
Finally, the associated scalar field is given by
\be
\label{scalprof}
\phi(r) = \phi_{0} + \frac{\Sigma}{2b} \log{\left[ \frac{r - M + b}{r-M-b}\right]} \, .
\ee
This  solution depends on three parameters: the mass $M$, the (rescaled) angular momentum $a$ and the scalar charge $\Sigma$. When the scalar charge vanishes, i.e $\Sigma =0$, the metric reduces to the Kerr geometry as expected. The effect of the massless scalar field is encoded in the function $\zeta(r,\theta)$.

The singularities of the solution can be tracked by computing the Ricci scalar which reads
\be
R = \frac{2\Sigma^2 }{\Delta \zeta \left( r^2 + a^2 \cos^2{\theta}\right)}  \, .
\ee
At constant angle $\theta$ when $\Delta \rightarrow 0$, we see that $R \sim \Sigma^2 r^{-2} \Delta^{-1-\Sigma^2/b^2}$.
Hence the hypersurface defined by $\Delta =0$ is singular when $\Sigma \neq 0$, which signals that the Kerr outer horizon has been turned into a curvature singularity because of the presence of the scalar field. If $M \geqslant a$, the location of this singular hypersurface is at
\be
r_{+} = M + M \sqrt{1 -  J^2} \leqslant 2M \, ,
\ee
where $J \equiv a /M$.  Consequently, this solution is only defined for $r \in \; ] r_{+}, + \infty [$ and describes the gravitational field of rotating compact object endowed with a scalar charge.

Setting $\Sigma=0$ and taking the non-rotating limit $a\rightarrow 0$, the geometry reduces to a deformed Schwarzschild metric which corresponds to a sub-class of the Zipoy-Voorhees (ZV) metric with scalar source which was first derived in \cite{Fisher:1948yn, Janis:1970kn}. See \cite{Astorino:2014mda, Turimov:2018guy, Chauvineau:2018zjy} for more recent generalizations. The vacuum ZV metric represents the simplest static and axi-symmetric vacuum solution of GR \cite{Voorhees:1971wh}. See \cite{Toktarbay:2015lua, Quevedo:2013qza} for details. We shall now investigate the disformal transformation of this Einstein-Scalar exact solution. 

\subsection{Disformed generalized Kerr solution}

Performing the same constant disformal transformation as the one used in the core of the paper, we obtain a new exact solution in  DHOST theories whose metric still takes the form 
\begin{align}
ds^2 = g_{\mu\nu} dx^{\mu} dx^{\nu} =   - f \left( dt - \omega d\psi \right)^2 + \frac{h_{ij}}{f} dx^i dx^j \, ,
\end{align}
with the same scalar profile (\ref{scalprof}).
The only modification, compared to the previous undeformed solution, shows up in the $g_{rr}$ components.  
Indeed, the spatial metric $h_{ij}$ reads now
\be
h_{ij} dx^i dx^j = H \left( G dr^2 + \Delta \, d\theta^2 \right) + \Delta \sin^2{\theta}\, d\psi^2
\ee
with
\begin{align}
H  =  \frac{\rho f }{\Delta}\left( 1 + \frac{b^2}{\Delta} \sin^2{\theta}\right)^{- \Sigma^2 /b^2} \;, \qquad 
 G = 1 -  \frac{B_{0} f}{H} \left( \phi' \right)^2 \, .
\end{align}
where a prime denotes derivative w.r.t the radial coordinate $r$.
As expected, the new solution has now four parameters: the mass $M$, the rescaled angular momentum $a$, the scalar charge $\Sigma$ and the disformal parameter $B_0$. We shall again only consider sufficiently small values of this parameter which ensures that the deformation function $G$ does not generate any singularity.
We can now investigate the properties of this new exact DHOST solution.

\medskip

First, we study the static limit of the disformed geometry. When the angular momentum vanishes, i.e.\ $a\rightarrow 0$, the DHOST solution reduces to the following metric
\begin{align}
ds^2 & = - \left( 1 - \frac{2M}{r}\right) dt^2 + r (r-2M) \sin^2{\theta} \, d\psi^2 \nonumber \\
& + \left( 1 - \frac{2M}{r}\right)^{-1 + \Sigma^2/M^2} \left( 1 - \frac{2M}{r} + \frac{M^2}{r^2} \sin^2{\theta}\right)^{-\Sigma^2/M^2} \left( G \, dr^2 + r(r- 2M) \, d\theta^2 \right)
\end{align}
while the scalar profile (\ref{scalprof}) reduces to
\be
\label{scal}
\phi(r) = \phi_{0} - \frac{\Sigma}{2M} \log{\left[ 1-\frac{M }{r}\right]} \, .
\ee
The explicit form of the function $G$ which contains the disformal parameter is given by
\be
\label{g}
G(r) = 1 - B_{0} \left( 1 - \frac{2M}{r}\right)^{1 + \Sigma^2/M^2} \left( 1 - \frac{2M}{r} + \frac{M^2}{r^2} \sin^2{\theta}\right)^{-\Sigma^2/M^2} \left( \phi' \right)^2 \, .
\ee
Interestingly, this axisymmetric but static solution provides a DHOST generalization of the well known vacuum $q$-metric of GR \cite{Quevedo:2013qza} and its recent extension with a scalar source presented in \cite{Turimov:2018guy}.
When the scalar charge vanishes, i.e.\ $\Sigma = 0$, the solution reduces to the Schwarzschild metric. As such, this new solution provides an new static axi-symmetric deformation of the Schwarzschild geometry with a quadrupole momentum in  DHOST theories. Let us also mention that, if one computes the gradient of the scalar field which enters in (\ref{scal}), we show that $G(r) \sim 1$ when $r\rightarrow +\infty$, and there is therefore no conical defect.

\medskip

We  now  investigate the causal structure. As the disformal parameter appears only in the $g_{rr}$ and $g_{\theta\theta}$ components of the metric, one can easily compute principal null directions given by
\begin{align}
\ell_{+}^{\alpha}\partial_{\alpha} & = \frac{\Delta}{\rho} \left( 
 \frac{r^2 + a^2}{\Delta} dt + \frac{dr}{G^{1/2}\zeta^{1/2}} +
 \frac{a}{\Delta} d\psi  \right) \, ,\\
\ell_{-}^{\alpha}\partial_{\alpha} & = 
  \frac{r^2 + a^2}{\Delta} dt - \frac{dr}{G^{1/2}\zeta^{1/2}} + 
 \frac{a}{\Delta} d\psi  \, ,
\end{align}
such that $g_{\alpha\beta}\ell_{\pm}^{\alpha} \ell^{\alpha}_{\pm} = 0 $ and $g_{\alpha\beta} \ell_{+}^{\alpha} \ell^{\beta}_{-} =-2$. 
The associated expansions  are given by
\begin{align}
\Theta_{+} & = \frac{ r \left( r-2M\right) +a^2 }{4  \sqrt{G \zeta}} \ \left( \frac{4 r}{\rho} +\frac{\zeta'}{\zeta} \right) \, ,\\
\Theta_{-} & = - \frac{1}{4\sqrt{G \zeta}} \left(  \frac{4 r}{\rho} +\frac{\zeta'}{\zeta} \right) \, .
\end{align}
Therefore, the product of the expansions reads
\be
\Theta_{+} \Theta_{-} =- \frac{ r \left( r-2M\right) +a^2 }{4 |G|| \zeta|} \left( \frac{4 r}{\rho} +\frac{\zeta'}{\zeta} \right)^2 \, .
\ee
The effect of the disformal transformation appears through the function $G$ which depends explicitly on $B_{0}$. We observe that this function appears either as a square root in the individual expansion or as an absolute value in the product $\Theta_{+} \Theta_{-} $. Therefore the disformal transformation cannot change the global sign of this quantity and the causal structure remains the same as the GR one in the DHOST frame. The new solution is horizonless geometry and it describes thus a rotating naked singularity.

\providecommand{\href}[2]{#2}\begingroup\raggedright\endgroup

\end{document}